\pdfoutput=1

\documentclass[11pt]{article}

\usepackage[hyperref]{acl2021}
\usepackage{times}
\usepackage{latexsym}

\usepackage{microtype}
\usepackage{multirow}

\usepackage{bbm}

\usepackage[toc,page]{appendix}
\usepackage{hyperref}
\usepackage{url}
\usepackage[neverdecrease]{paralist}
\usepackage{graphicx}
\usepackage{booktabs}
\usepackage{graphics}
\usepackage{tabularx}
\usepackage{xspace}
\usepackage{graphicx}
\usepackage{booktabs}
\usepackage{mdwlist}
\usepackage{multirow}
\usepackage{amsmath}
\usepackage{mathtools}
\usepackage{calc}
\usepackage[neverdecrease]{paralist}
\usepackage[ruled,noend]{algorithm2e}
\usepackage{booktabs}
\usepackage{diagbox}
\usepackage{colortbl}
\usepackage{subcaption}
\usepackage{makecell}
\usepackage{verbatim}
\usepackage{amsmath}
\usepackage{color,soul}
\aclfinalcopy

\usepackage{graphicx}

\newcommand{\heart}{\ensuremath\heartsuit}

\usepackage{fdsymbol}

\DeclareSymbolFont{extraup}{U}{zavm}{m}{n}
\DeclareMathSymbol{\varheart}{\mathalpha}{extraup}{86}
\DeclareMathSymbol{\vardiamond}{\mathalpha}{extraup}{87}

\newcommand{\numLangs}{16 }
\newcommand{\numProxyLangs}{11 }
\newcommand{\numTrainLangs}{27 }
\newcommand{\numIntents}{14 }

\title{Global Readiness of Language Technology for Healthcare: What would it Take to Combat the Next Pandemic?}

\author{$^\heart$Ishani Mondal\thanks{\hspace{0.1cm} Equal contribution}, $^\heart$Kabir Ahuja\footnotemark[1], $^\heart$Mohit Jain, $^\diamondsuit$Jacki O'Neil,  $^\heart$Kalika Bali,
$^\heart$\textbf{Monojit Choudhury} \\ $^\heart$Microsoft Research Labs, Bangalore, India \\ $^\diamondsuit$Microsoft Africa Research Institute, Nairobi, Kenya \\
\fontfamily{qcr}\selectfont{\{t-imonda, t-kabirahuja, kalikab, monojitc\}@microsoft.com} \\}

\begin{document}
\maketitle
\begin{abstract}

The COVID-19 pandemic has brought out both the best and worst of language technology (LT). On one hand, conversational agents for information dissemination and basic diagnosis have seen widespread use, and arguably, had an important role in combating the pandemic. On the other hand, it has also become clear that such technologies are readily available for a handful of languages, and the vast majority of the global south is completely bereft of these benefits.
What is the state of LT, especially conversational agents, for healthcare across the world's languages? And, what would it take to ensure global readiness of LT before the next pandemic? In this paper, we try to answer these questions through survey of existing literature and resources, as well as through a rapid chatbot building exercise for 15 Asian and African languages with varying amount of resource-availability.
The study confirms the pitiful state of LT even for languages with large speaker bases, such as Sinhala and Hausa, and identifies the gaps that could help us prioritize research and investment strategies in LT for healthcare.

\end{abstract}

\section{Introduction}
The world witnessed one of the worst pandemics in early 2020, COVID-19, infecting over 250 million people globally.
Scientists and technologists from various fields joined hands, lending support to deal with this global crisis. Language Technology (LT) played a crucial role in combating the pandemic  through the development of healthcare chatbots that facilitate information dissemination~\cite{li-etal-2020-jennifer, fi12070109} and early disease screening~\cite{10.1093/jamia/ocaa130,Martin2020}. Nevertheless, today practically useful chatbots and other benefits of LT are available only in a handful of languages~\cite{joshi-etal-2020-state}. Despite impressive gains made by the Massively Multilingual Transformer based Language Models (MMLM) \cite{devlin-etal-2019-bert, Lample2019CrosslingualLM, 
aharoni-etal-2019-massively, conneau-etal-2020-unsupervised, xue-etal-2021-mt5} on standard NLP benchmark tasks \cite{Pan2017, Conneau2018xnli, Yang2019paws-x, hu2020xtreme}, the real-world implications of such advancements remain largely unexplored. \citet{joshi-etal-2020-state} has highlighted such a disparity and classified the languages of world into six classes based on their resource-availability, where Class 5 represents the most resource-rich languages for whom benefits of LT are readily available; and class 0 denotes the most under-resourced languages. 
In this paper, we ask the following two questions: (1) Today, in which languages can we build practically useful LT systems, especially chatbots, that could help us combat a pandemic? (2) How should we prioritize research and resource building investments so that LT is globally ready before the next pandemic?

In order to answer these questions, we review the existing literature and resources on COVID-19 chatbots, and classify them based on the languages they support and the solutions they provide. Quite unsurprisingly, the survey reveals a strong disparity in LT solutions between resource-rich and resource-poor languages. In order to quantify this gap and measure the pandemic-readiness of various languages today, we select 15 Asian and African languages (except English) with various degrees of resource-availability, and attempt to build COVID-19 FAQ bots for them. Since building an end-to-end chatbot is a substantial engineering effort, we scope the problem down to building an intent classifier for these languages, which forms the core of the Natural Language Understanding (NLU) unit. We also experiment with entity recognition for a subset of these languages. 

Our study shows that despite using the best available commercial multilingual chatbot frameworks  (e.g., Google Dialogflow\footnote{\url{https://cloud.google.com/dialogflow}}, Microsoft Bot Framework (MS Bot)\footnote{\url{https://dev.botframework.com/}}), advanced Machine Translation (MT) systems\footnote{\url{https://translate.google.co.in/, https://www.bing.com/translator}}, and state-of-the-art massively multilingual language models (mBERT \cite{devlin-etal-2019-bert} and XLM-R \cite{conneau-etal-2020-unsupervised}), there is a 20-30\% drop in performance for class 0 - 2 languages as compared to English. The drop is large for all the African languages (e.g., Hausa and Somali) and some of the Asian languages (e.g., Marathi and Sinhala). Note that our experiments were limited to languages which are supported by at least one of the chatbot frameworks, MT systems or MMLMs. There are thousands of other languages which are supported by none.

We extrapolate our findings at global scale and construct a global LT readiness map for pandemic-response and healthcare. Based on this map as well as error analysis of the chatbot experiments, we identify a set of research problems as well as resource-prioritization strategies which we believe are key to ensure global LT readiness before the next pandemic. 

The rest of the paper is organized as follows: Sec ~\ref{sec:survey} presents the literature survey on LT response to COVID-19, specifically focusing on chatbots built for the pandemic. Sec ~\ref{sec:botbuilding} describes the chatbot building experiments, where in ~\ref{subsec:lang_selection} we motivate the choice of languages for the experiment, in ~\ref{sec:intent} and ~\ref{sec:entity} we discuss the intent and entity detection experiments respectively. In Sec ~\ref{sec:readiness} we present the global LT readiness map and in Sec \ref{sec:reco}, we conclude with our recommendations.

\begin{figure*}[t]
\centering

\begin{subfigure}[t]{0.48\textwidth}
    \includegraphics[width=0.9\textwidth]{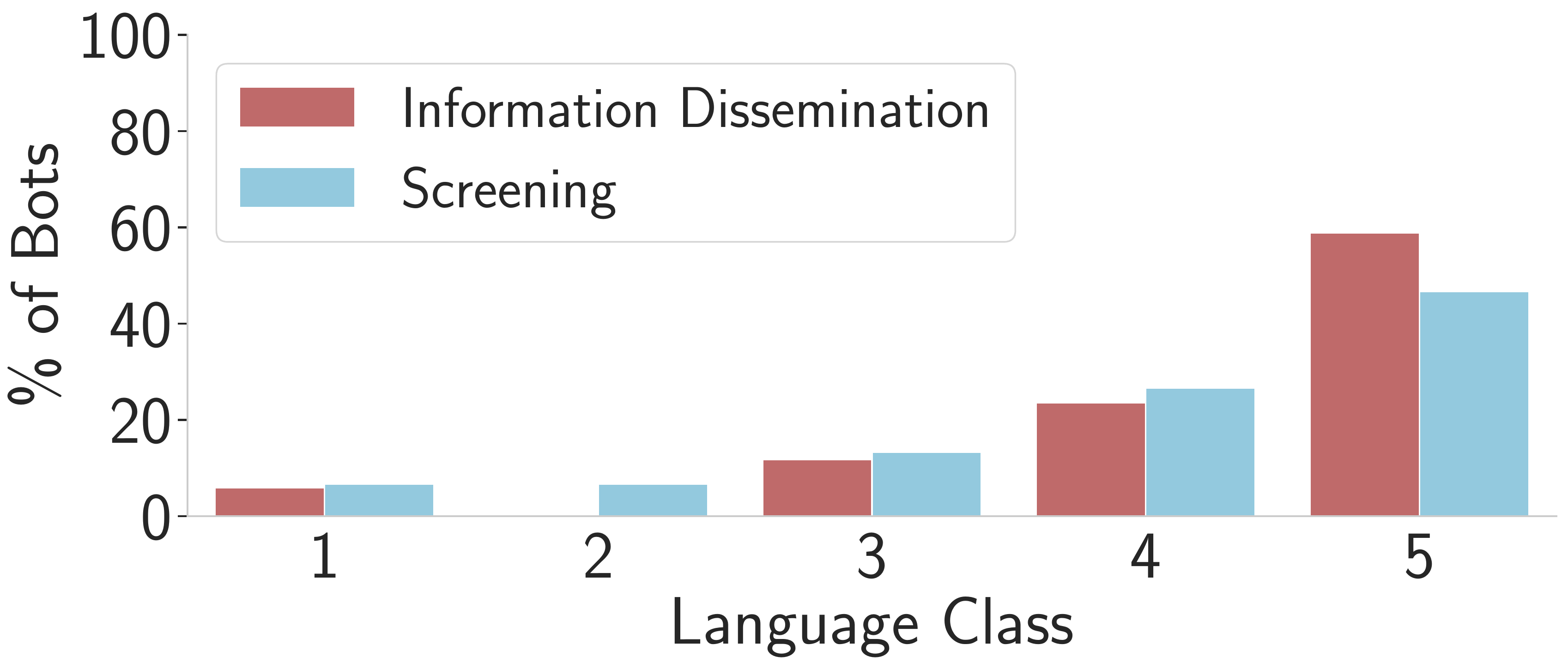}
    \caption{Percentage of papers covering the development of COVID-19 chatbots per language class}
    \label{fig:num_papers}
\end{subfigure}%
\hfill
\begin{subfigure}[t]{0.48\textwidth}
    \includegraphics[width=0.9\textwidth]{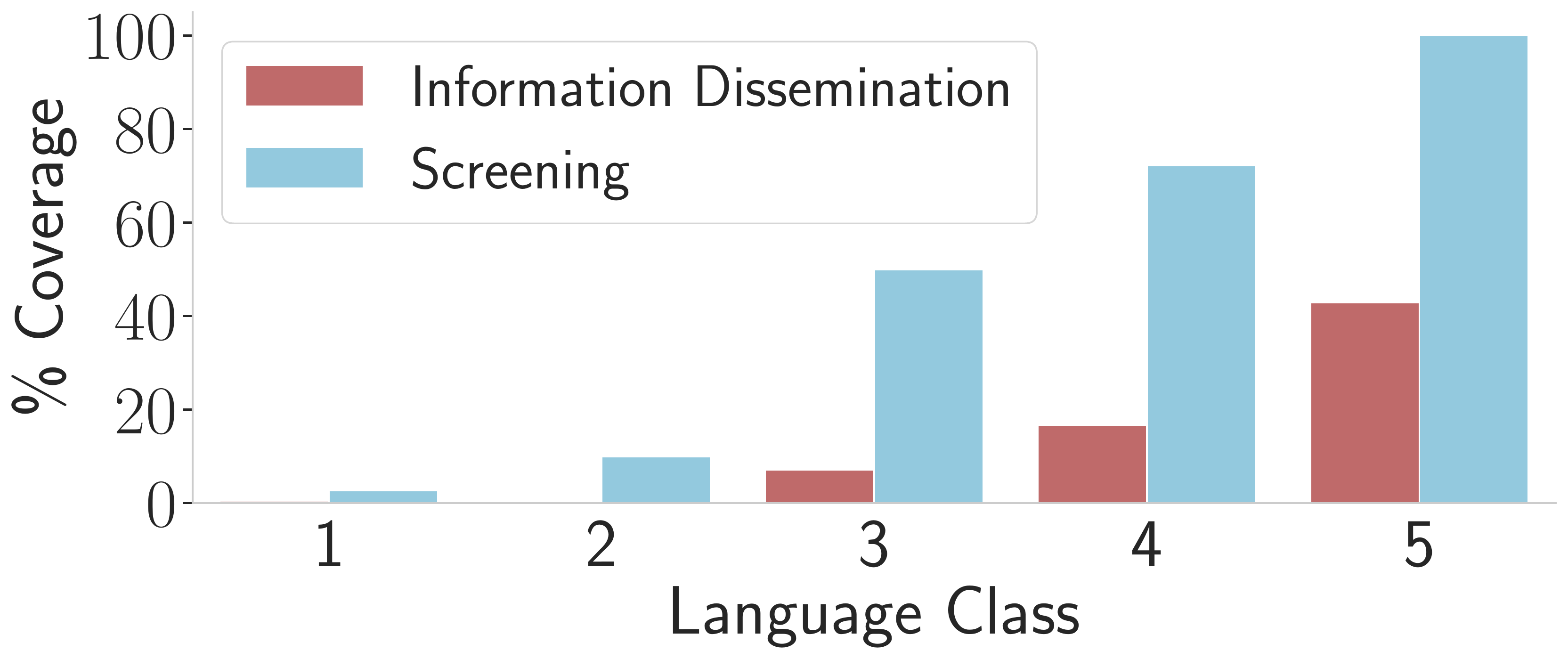}
    \caption{Percentage of languages of each class for which at-least one paper addressed the development of a chatbot.}
    \label{fig:papers_coverage}
\end{subfigure}

\begin{subfigure}[t]{0.48\textwidth}
    \includegraphics[width=0.9\textwidth]{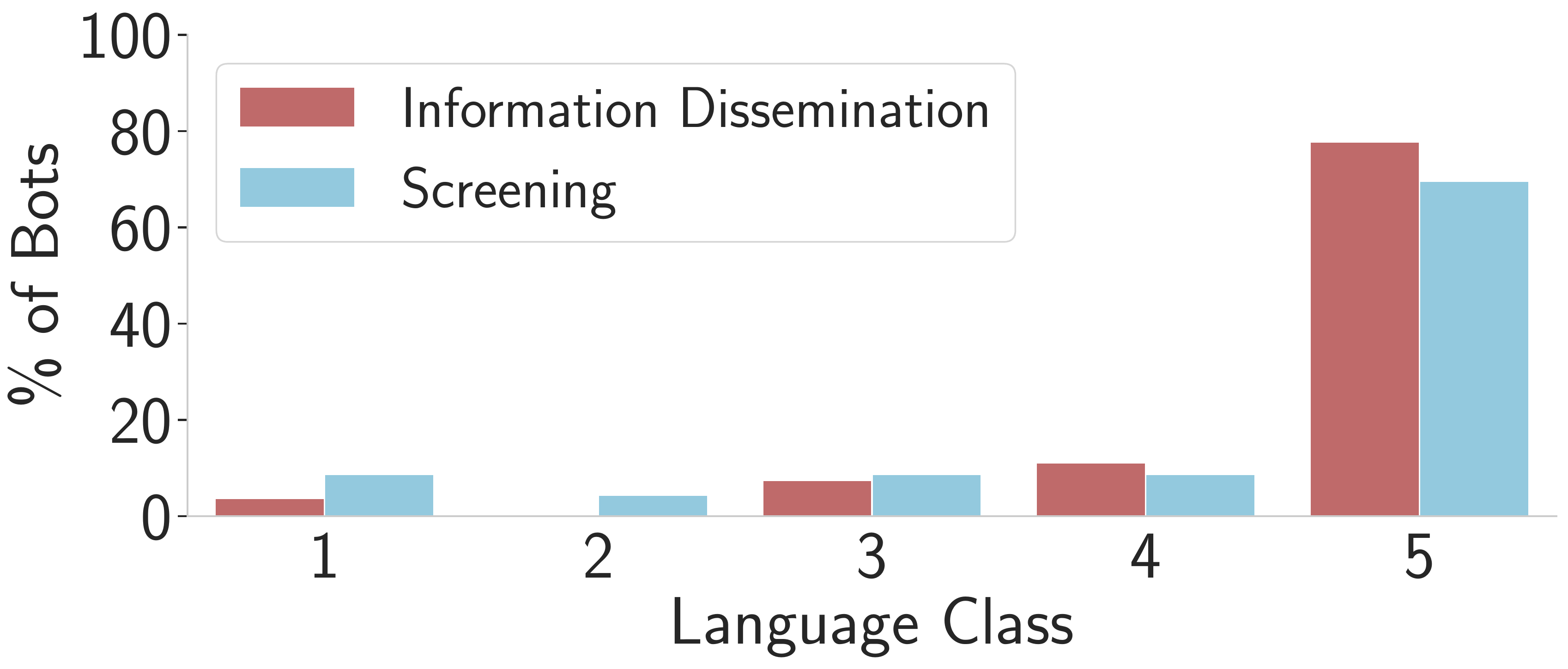}
    \caption{Percentage of bots deployed by governmental agencies , provider organizations etc. per language class, to combat the pandemic situation.}
    \label{fig:num_deps}
\end{subfigure}%
\hfill
\begin{subfigure}[t]{0.48\textwidth}
    \includegraphics[width=0.9\textwidth]{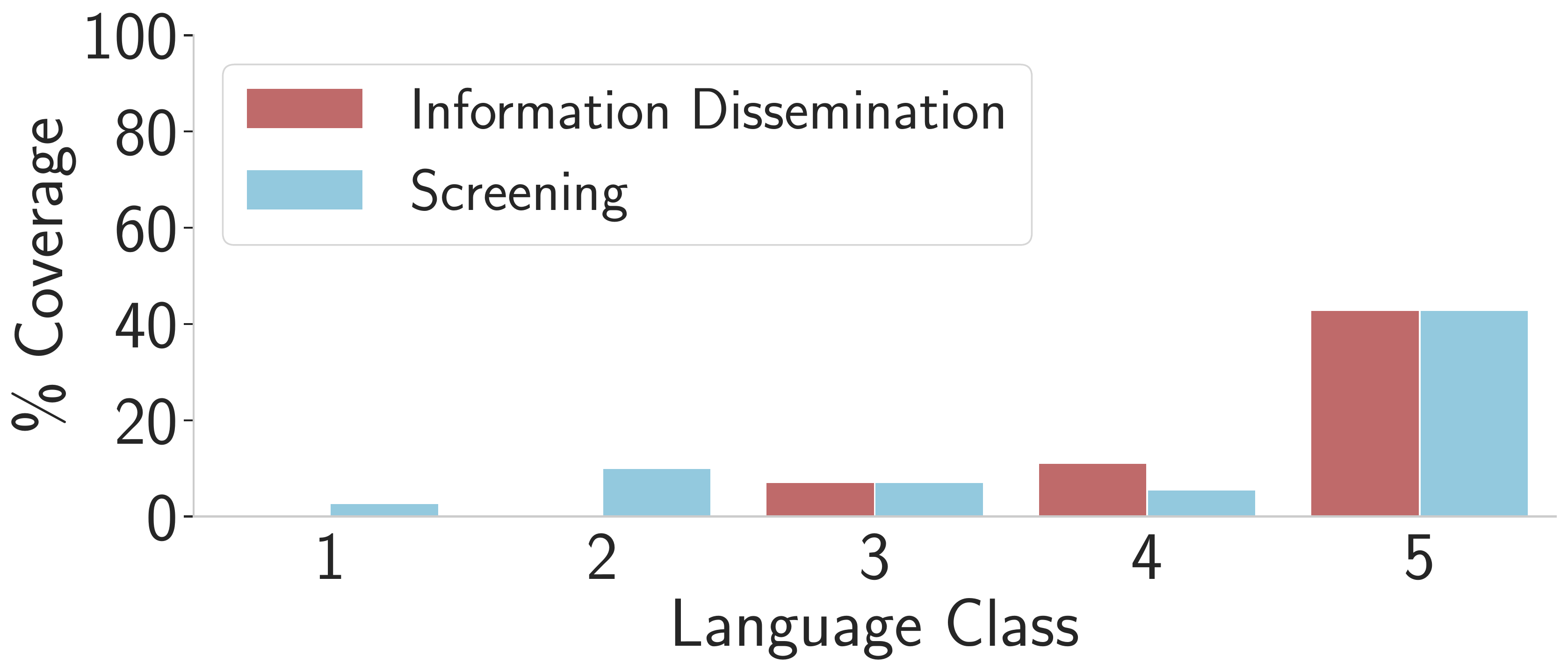}
    \caption{Coverage of languages across different classes in the bots deployed by governmental agencies , provider organizations etc.}
    \label{fig:deps_coverage}
\end{subfigure}
\caption{The number of bots developed (for both research and public deployment use cases) for different language classes and their coverage.}
\label{fig:mnist}
\end{figure*}

\section{Literature Survey}
\label{sec:survey}
In the recent years, NLP for Healthcare has witnessed a major uptake and an impressive volume of work has significantly pushed the research forward by developing sophisticated domain-specific language models \cite{alsentzer-etal-2019-publicly, Lee2020BioBERTAP, Ji2021MentalBERTPA}. 
These models have been adopted to serve different axes of healthcare such as patient provider communication \cite{pmlr-v126-min20a, pmlr-v126-si20a},
information dissemination \cite{fi12070109}, and
self-care management and therapy \cite{info:doi/10.2196/10148, 8784055,  info:doi/10.2196/12231, Kamita2019}. 
The role of healthcare chatbots becomes crucial along all these axes because of the recent adoption of telehealth technology services \cite{10.1145/3476064}. 
 
Chatbots have received a considerable interest during the recent COVID-19 pandemic. Due to the worldwide spread and severity of the virus and subsequent global response, we believe that the study of COVID-19 chatbots can provide us an accurate picture of the global-readiness of LT. We, therefore, surveyed COVID-19 chatbots that are mentioned in the literature and/or deployed in the real-world.\footnote{Besides healthcare, NLP has also proved beneficial in providing aid during natural disasters like earthquakes and floods \cite{lewis-2010-haitian, 10.1145/2806416.2806485, 8944883, w11020234, 8715653}. 
\citet{strassel-tracey-2016-lorelei} leveraged existing language technologies for resource-poor languages to combat natural disasters. Though this study is limited to pandemic readiness, we believe the state of LT for disaster-readiness across the globe would be very similar.} 

\subsection{Use-Cases and Technological Support}
~\label{usecase}
From the survey, two primary use-cases of COVID-19 chatbots emerge -- (1) information dissemination: answering pandemic-related questions asked by the users \cite{li-etal-2020-jennifer, covdeep, prasannan-etal-2020-chatbot, Mehfooz2020MedicalCF, Trang2021EnhancingRN}, and (2) symptom-screening: assessing risk factors associated with the symptoms provided by the user for quick diagnosis \cite{Ferreira2020ANAAB, Martin2020AnAI, Judson2020ImplementationOA, 10.1145/3457784.3457788}.
Existing commercial frameworks such as DialogFlow, Watson Assistant and MS Bot have been used primarily for building a majority of these chatbots \cite{li-etal-2020-jennifer, 9532611}.
However, open-source bot frameworks like Rasa \cite{10.1145/3457784.3457788, vietnamese} have also been gaining traction in the community.
The inbuilt NLU engines supported by these frameworks makes chatbot development easy, hence there is a significant uptake in utilizing these to develop new chatbots.
Pre-trained LMs were also leveraged for COVID symptom identification \cite{10.1145/3388440.3412413} and question answering \cite{park2020classification}.

\subsection{Language Diversity}
~\label{multilingual}
Which languages are supported by these COVID bots? Of the 20 COVID-related bots mentioned in the existing literature and 34 others deployed by different countries to combat the pandemic, 26 ($\approx 50$\%) are exclusively for English, followed by German having 10 deployed bots. In Figure \ref{fig:mnist}, we show the distribution of the chatbots by the language classes defined in \citet{joshi-etal-2020-state}. As expected, for all the cases, we observe that chatbots were available primarily and almost exclusively for languages in class 5. We observe a slightly higher presence of class 4 and 3 languages in research papers on COVID chatbots (Fig.~\ref{fig:num_papers}). For instance, there are three research papers each for Hindi and Vietnamese, both class 4 languages. To the best of our knowledge, we could not find any publication or deployed bot for class 0 languages.

This skew is more prominent when we consider the coverage of languages of different classes, i.e., the fraction of languages in each class for which at least one COVID-19 conversation system was developed (Figure \ref{fig:papers_coverage} and \ref{fig:deps_coverage}).
This lack of attention to a large number of languages has also been highlighted by \citet{anastasopoulos-etal-2020-tico} who strongly advocated for the development of language resources for improving access to COVID-19 related information in 26 lesser-resourced languages, particularly from Africa and South  and South-East Asia.

\section{Rapid Chatbot Building Exercise}
\label{sec:botbuilding}
How quickly can one build a pandemic response chatbot in a language based on the best publicly available systems? 
In order to answer this question we have to understand the pandemic-readiness of various languages. To do this, we made an attempt to build chatbots for answering frequently asked questions about COVID-19 using Google Dialogflow,  Microsoft Bot Framework (MS Bot), as well as two of the most popular Massively Multilingual Language Models (MMLM) -- mBERT and XLM-R. Since building an end-to-end chatbot is complex, we chose to conduct rapid prototyping experiments  for intent recognition in \numLangs languages, and entity recognition in 3 languages.

\subsection{Language Selection Criteria}
\label{subsec:lang_selection}
For our experiments, we chose a few languages from each language class (defined by \citet{joshi-etal-2020-state}) such that at least one language per class is supported by either of the two commercial chatbot frameworks, leading to the following set: \emph{English, Chinese} from Class 5, \emph{Hindi, Korean} from Class 4,  \emph{Bengali, Malay} from Class 3,  \emph{Swahili, Hausa, Marathi, Amharic, Zulu} from Class 2, \emph{Assamese, Gujarati, Kikuyu, Somali} from Class 1, and \emph{Sinhala} from Class 0.

\subsection{Intent Recognition}

\label{sec:intent}
Intent Recognition is an essential component of conversational systems. Given a user query, the task is to classify it into one of the pre-defined intent categories 
\cite{braun-etal-2017-evaluating}.

\subsubsection{Dataset Creation and Characteristics}
For training and evaluation, we curate a set of 147 queries categorized into one of the \numIntents intents:
1) \texttt{Airborne} (how COVID spreads by air), 2) \texttt{ClarifyCovid} (difference between COVID and other diseases), 3) \texttt{Country} (country-wise infection statistics), 4) \texttt{CovidTwice} (possibility of reinfection), 5) \texttt{ExplainSymptom} (COVID symptoms) 6) \texttt{Incubation} (how many days of incubation required), 7) \texttt{Length} (longevity of infection), 8) \texttt{Mask} (ways of wearing mask), 9) \texttt{Protection} (ways to protect against infection), 10) \texttt{Quarantine} (quarantine requirement of US), 11) \texttt{Spread} (how COVID spreads), 12) \texttt{Testing} (available COVID tests), 13) \texttt{Medication} (about drugs to protect from COVID), and 14) \texttt{Treatment} (about treatments or therapies related to COVID). Examples and definitions of each intent are present in ~\ref{appendix:intentDesc}

We refer to the FAQs provided by the UN~\cite{UNDoc} and user queries in the dataset released by \citet{anastasopoulos-etal-2020-tico}, to identify the \numIntents types of questions that a user may ask.
We manually paraphrase the questions to generate queries (Mean = 10.5, S.D. = 4.36 queries per intent) for each intent in English.
Two annotators with native English proficiency independently classified these queries;
the inter-annotator agreement ($\kappa$) was 0.89.
We asked a few native speakers of each of the selected languages to translate these 147 queries manually.
The dataset is split into train and test sets, using a stratified split over the intents, giving a total of 76 and 71 queries in train and test set respectively.

\subsubsection{Strategies of Developing Chatbots}
We consider three training and inference strategies, emulating the possible scenarios for developing such chatbots in practical settings (details in ~\ref{appendix:bot}).
\noindent{\textbf{Train on English Data}}:
In this strategy, we develop our bots by training them on the English queries, and evaluate the intent detection performance in different languages by automatically translating the test queries into English (e.g., similar to \citet{gupta2021truthbot}). 

\noindent{\textbf{Train on MT Translations}}: Here we build target language intent classifier models from training data in different languages, which is obtained by automatically translating the English training data. The classifier is then tested on the manually translated test data in the corresponding target language. A similar method was adopted by \citet{balahur-turchi-2012-multilingual} for sentiment analysis.

\noindent{\textbf{Train on Manual Translations}}: In this setup, we use the manual translations of the English training dataset to train our bots in different languages. Like the previous setup,
here again we use the manually translated data to evaluate the intent detection performance of the developed chatbots. Jennifer Bot \cite{li-etal-2020-jennifer} used a similar setup to extend their English bot to Spanish. Note that this is the most expensive setup in terms of data creation cost.

\begin{table*}[t]
\footnotesize
\centering
\begin{tabular}{l l rr rr rr rr}
\toprule
&& \multicolumn{2}{r}{Train on English Data} & \multicolumn{2}{r}{Train on MT Translations} & \multicolumn{4}{r}{Train on Manual Translations}\\
\cmidrule(r){3-4}
\cmidrule(r){5-6}
\cmidrule(r){7-10}
Class & Languages & DF & LUIS & DF & LUIS & DF & LUIS & mBERT & XLM-R\\
\midrule
\multirow{1}{2em}{5}& Chinese & 0.60
 & 5.00
 & $17.50^\dagger$ & $18.40^\dagger$ & $0.04$ & $(5.20)$& $(5.63)$
 & $(8.50)$\\
 \cmidrule(r){1-1}
\multirow{2}{2em}{4} & Hindi & 
$12.50$ & $0.05$ &  $16.25^\dagger$ & $25.01^\dagger$ & $13.02$ &
$10.50$ & $12.72^\dagger$ & $19.15^\dagger$\\
& Korean & $6.50$ & $13.71^\dagger$ & $31.20^\dagger$ & 11.55  
 &  $23.75^\dagger$ & $10.00$ & $5.63^\dagger$&$10.88^\dagger$\\
 \cmidrule(r){1-1}
\multirow{2}{2em}{3}&Bengali &
$20.50^\dagger$  & $13.15^\dagger$ &  $26.25^\dagger$ & $\times$ & 11.80 &
$\times$& $7.04^\dagger$ & $2.13^\dagger$\\
&Malay &
$21.24^\dagger$ & 11.87 &  $19.53^\dagger$ & $\times$ & $12.50$ &$\times$& 4.77
 & $14.89^\dagger$ \\
 \cmidrule(r){1-1}
\multirow{3}{2em}{2}&Swahili &
$28.08^\dagger$ & $18.00^\dagger$ & $\times$ &
$\times$ & $\times$ & $\times$ & $32.39^\dagger$ & $19.15^\dagger$  \\
&Hausa &
$40.97^\dagger$ & $34.00^\dagger$ & $\times$ &
$\times$ & $\times$ & $\times$ & $\times$ & $29.79^\dagger$ \\
&Marathi &
$21.23^\dagger$ & $14.00^\dagger$ & $28.08^\dagger$ & $28.7^\dagger$ & $16.25^\dagger$ &
$17.76^\dagger$& $16.90^\dagger$ & $29.79^\dagger$\\
&Amharic &
$43.06^\dagger$ & $34.82^\dagger$& $\times$ &
$\times$ & $\times$ & $\times$ & $\times$  &  $12.39^\dagger$ \\
&Zulu &
$30.56^\dagger$ & 11.28 & $\times$ & $\times$ & $\times$ &
$\times$ & $\times$  & $\times$  \\
 \cmidrule(r){1-1}
\multirow{2}{2em}{1}&Assamese &
$19.52^\dagger$ & $18.00^\dagger$ & $\times$ &
$\times$ & $\times$ & $\times$ &
$\times$ & $29.79^\dagger$\\
&Gujarati &
 15.55 & 10.03 & $\times$ & $22.88^\dagger$
 & $\times$ & $22.88^\dagger$  & 4.77 & $19.15^\dagger$\\
&Kikuyu* & $97.60^\dagger$
 & $76.87^\dagger$ & $\times$ &
$\times$ & $\times$ & $\times$ & $\times$ & $\times$ \\
&Somali &
$40.56^\dagger$ & $27.58^\dagger$ & $\times$ &
$\times$ & $\times$ & $\times$ & $25.35^\dagger$ & $\times$\\
 \cmidrule(r){1-1}
0&Sinhala &
$35.00^\dagger$ & $19.00^\dagger$  & $34.93^\dagger$ & $\times$ & 15.65 &
$\times$ & $61.97^\dagger$ & $19.15^\dagger$ \\
\bottomrule
\end{tabular}
\caption{$\delta_{l}$ for each language for the \textit{Intent Recognition} task using the three different strategies. $\times$ indicates that the framework does not support end-to-end chatbot development for that language. Drops that lead to accuracy below 67\% are marked by $\dagger$, indicating the case where the bot mis-recognizes 1 out of every 3 queries.  *Owing to non-availability of standard MT for Kikuyu, we used \textit{Safarini}\footnote{\url{https://play.google.com/store/apps/details?id=com.dictionary.safarini&hl=en_IN&gl=US}} app from Android playstore for translation. Note: The values mentioned in the parantheses indicate that we observe relative gain instead of drop.}
\label{tab:results}
\end{table*}

\begin{figure}
     \centering
     \includegraphics[width=0.48\textwidth]{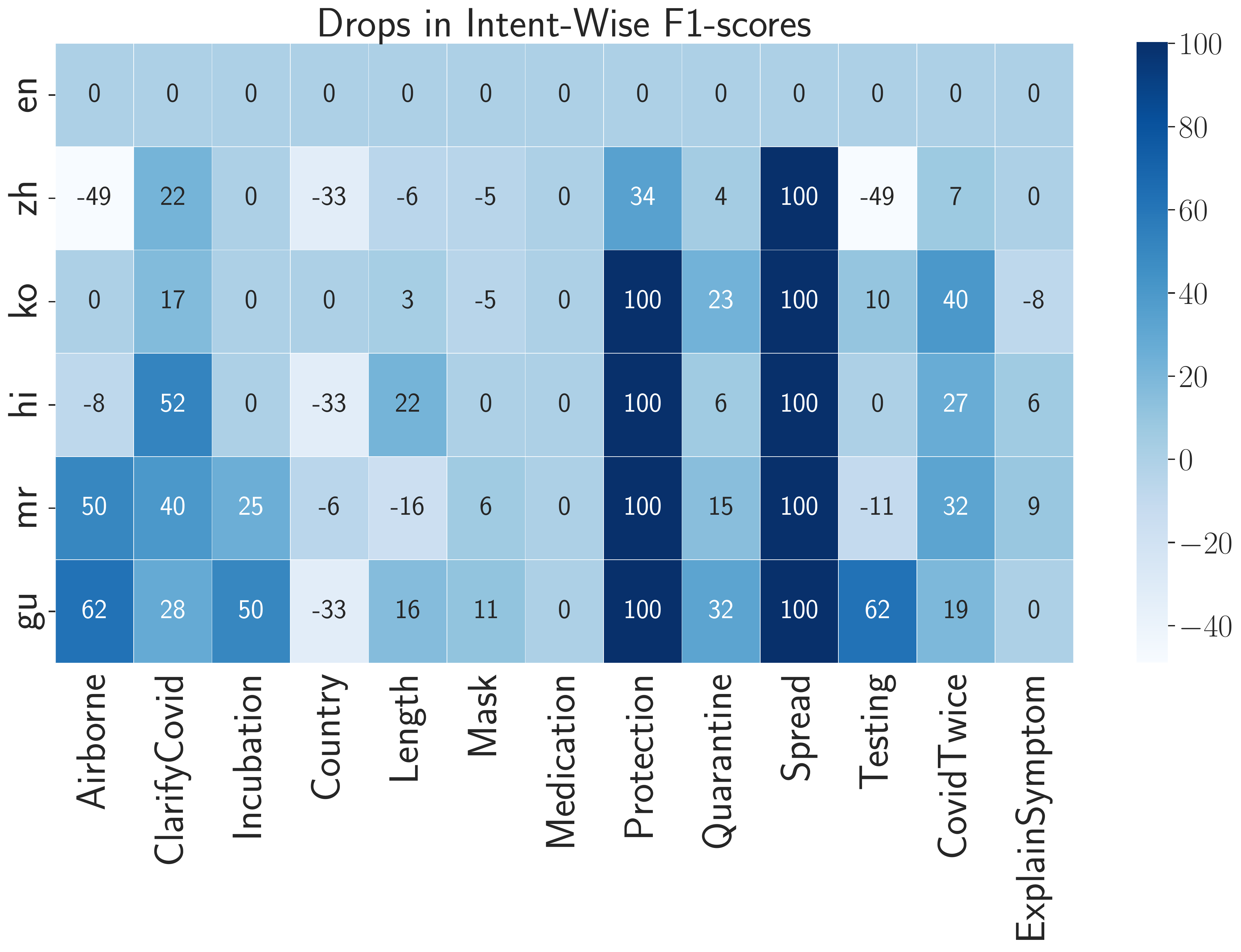}
     \caption{Relative drops (relative to English) in intent wise F1 scores for different languages in the \textit{Train on Manual Translations} setup (in LUIS). Negative values indicate increase in the scores relative to English.} 
     \label{fig:intent_f1}
\end{figure}

\subsubsection{Experimental Setup}

\noindent{\textbf{Commercial Frameworks}}: We use Google Dialogflow and MS Bot Framework to train and evaluate the FAQ bots in different languages. For Dialogflow we use the ES Console, and for 
MS Bot, we use Microsoft's Language Understanding Service (LUIS)\footnote{\url{https://www.luis.ai/}} framework.
Dialogflow and LUIS supports 7 and 6 out of our \numLangs selected languages, respectively. For the unsupported languages, we could only experiment with \textit{Train on English Data} setup.

\noindent{\textbf{Pre-trained MMLMs}}:
We evaluate two popular MMLMs, namely mBERT (\textit{bert-base-multilingual-cased}) and XLMR (\textit{xlm-roberta-base}), for our intent detection experiments. XLM-R supports all but Kikuyu, Somali and Zulu, while mBERT supports all but Amharic, Assamese, Hausa, Kikuyu and Zulu.
For these models, we only evaluate the \textit{Train on Manual Translations} setup.
We experiment with two different approaches for building intent classifiers with these models: i) Fine-tune the pre-trained MMLM as a classifier on the training set, and ii) Training an end-to-end classifier. 
Since our dataset is small, training an end-to-end classifier might be prone to overfitting, hence we use pre-trained embeddings to fit a k-nearest neighbors model as done in \citet{caron2021emerging}. We report the best scores out of these two setups for both MMLMs (details in ~\ref{appendix:mmlm}).

\noindent{\textbf{Evaluation}}:
We report the relative accuracy drop $\delta_l$ for each target language $l$ from English ($en$), defined as $(A_{en}-A_l)/(A_{en}) \times 100$, where $A_l$ is the accuracy of intent classification for $l$ on the held-out test set. Thus, lower the value of $\delta_l$, better is the state-of-the-art of LT for the language $l$.

\begin{table*}[t]
\footnotesize
\centering
\begin{tabular}{p{0.03\textwidth} p{0.21\textwidth} p{0.34\textwidth} p{0.31\textwidth}}
\toprule
\textbf{Lang}  & \textbf{Issue}  & \textbf{Actual Example}  & \textbf{Misclassified Translated Example}\\
\midrule
\textbf{Si} & \textbf{Terminology Mismatch}  
 & I have hay fever though too. & I also have \color{red}{gonorrhea}. \\
\textbf{Bn} & \textbf{Fluency} 
 & Is SARS-CoV-2 airborne? & Does SARS-CoV-2 \color{red}{sit in the air?}\\
\textbf{Hu} & \textbf{Relevance} 
 & I got the virus. How long does it go on for? & I {\color{red}{Nasami Cutar}}. How long will it take?\\
\textbf{Hu} & \textbf{Fluency, Terminology Mismatch} 
 & How long should I wear a mask? & How long will I \color{red}{impose sanctions}?\\
\textbf{Hu} & \textbf{Terminology Mismatch} 
& Is it healthy to wear a mask during swimming? & Is it safe, can I wear \color{red}{fascist sanctions} when I swim?\\
\bottomrule
\end{tabular}
\caption{Excerpts of test instances showing bottlenecks of MT systems in the \emph{Train on English Data} setup.
\label{tab:examples}
}
\end{table*}

\subsubsection{Results and Analysis}
Table \ref{tab:results} presents the intent classification results. While the relative drop $\delta_l$ is reported, we also mark the values with a $\dagger$ wherein the {\em absolute} accuracy, $A_l$ falls below 67\% ($A_{UX}$). Such a classifier is not useful for real-world deployment as it will misclassify every third query. 
As expected, we observe high $\delta_l$ for languages belonging to class 3 or lower, with most of the accuracies below $A_{UX}$.

\noindent{\textbf{Comparison across the three setups}}:
We observe that for classes 4 and 5, \textit{Translate on English Data} performs at par or even better than the most expensive \textit{Train on Manual Translations} setup.
This may be because the MT translations from these languages to English is highly accurate.
On the other hand, for languages belonging to class 3 or lower, \textit{Train on Manual Translations} led to better performance, arguably due to poorer performance of the MT system.
Unfortunately, the \textit{Train on Manual Translations} method is the most expensive in terms of data curation cost, hence may be the hardest to implement in the midst of a pandemic.
The problem become worse because
a majority of class 3 and lower languages are not supported by current chatbot frameworks. 
Even when supported,
their performance is below the acceptable limit (e.g., Marathi, Gujarati).
One of the reasons is 
the difficulty in correctly identifying technical intents like \texttt{Airborne} and \texttt{Incubation} in such low-resource languages (Figure \ref{fig:intent_f1}).

Since a few of these low-resource languages are present in the pre-training dataset of mBERT and XLM-R, we can evaluate them for \textit{Train on Manual Translations}.
There is a similar pattern in accuracy drop for MMLMs, however the accuracy begin to fall below the acceptable limit (67\%) from class 4 languages onward. There is a remarkable drop in mBERT's accuracy for Sinhala (class 0).
In general, we find mBERT to outperform XLM-R, except for Swahili and Sinhala. 
This may be due to the higher representation of these languages in the pre-training corpus of XLM-R (CommonCrawl Corpus).
This strongly indicates the importance of the pre-training dataset size for developing LT, both in terms of absolute size as well as relative size to other languages
\cite{wu-dredze-2020-languages}. 
As expected, the performance in \textit{Train on MT Translations} setup is the worst among the three; except for Korean in LUIS, all values lie below $A_{UX}$, which could be a compounded effect of poor translation quality and inferior NLU solutions.

To conclude, all languages in class 3-5 had at least one solution yielding an acceptable accuracy, while all languages in class 0-2, except Gujarati, Sinhala and Zulu, had no acceptable solution.

\noindent{\textbf{Lost in Translation}}: Table \ref{tab:examples} shows the intent misclassification errors due to the errors in MT translations. The manual translation in the target language correspond to the `Actual Example' in English, and the phrase translated back to English for the \textit{Train on English Data} setup is reported under the `Misclassified Translated Example.' We categorize the translation errors as Terminology Mismatch, Fluency and Relevance \cite{li-etal-2020-jennifer}.
We find that domain-specific terms often get translated incorrectly into English (\ref{tab:examples}).
In a few cases, the translations result in unnatural queries resulting in loss of fluency, such as \textit{Does SARS-CoV-2 sit in the air?}. 
All these factors lead to poor performance of \textit{Train on English Data} setup for low-resource languages. We find that \textit{Terminology Mismatch} is the most common issue affecting the performance.
Interestingly, technical terms like \textit{incubation} does not exist in a few of our target languages, hence the manually written test queries in these languages just had the English term written in that language's script. In such cases, we found lesser performance drop compared to languages when equivalent vocabulary exists in the target language. 
E.g., high drops in F1-score for intents like ``\texttt{Quarantine}" and ``\texttt{Incubation}" in Hausa (76\%, 100\% respectively) and Amharic (56\%, 100\%) justify this, whereas for Zulu, where the human translator used English terms in their queries resulted in much lower drop in F1 scores (20\%, 0\%).

See appendix for the intent-wise F1 scores for different languages.

\noindent
\textbf{Implications}: 
Based on our experimental results, we wish to explore how to prioritize the resource-investment strategies to push the state of current LT forward.
Resource-poor languages mostly under-perform across all the three set-ups, so then {\em should we invest more towards developing better translation systems or focus more on improving the current NLU solutions for different languages?}
We observe that a good quality translation system can support building bots from scratch in a new language, and often performs on-par with the \textit{Train on Manual Translations} setup for high-resource languages (e.g., Korean, Hindi) and sometimes even for low-resource languages (e.g., Gujarati).
Building bots from scratch in a new language is resource-intensive, requiring rapid prototyping, which may be infeasible during a crisis. Therefore, a generic way to ensure pandemic-readiness in a language is by ensuring reasonably accurate MT systems similar to that for class 4 languages. 
Improving representation of low-resource languages in the pre-training datasets of existing multilingual models (specifically on domain-specific corpora as done by \cite{10.1145/3458754, Zhang2020MultiStagePF}) is yet another way to ensure preparedness.

\begin{figure*}
    \centering
    \fbox{\includegraphics[width=\textwidth]{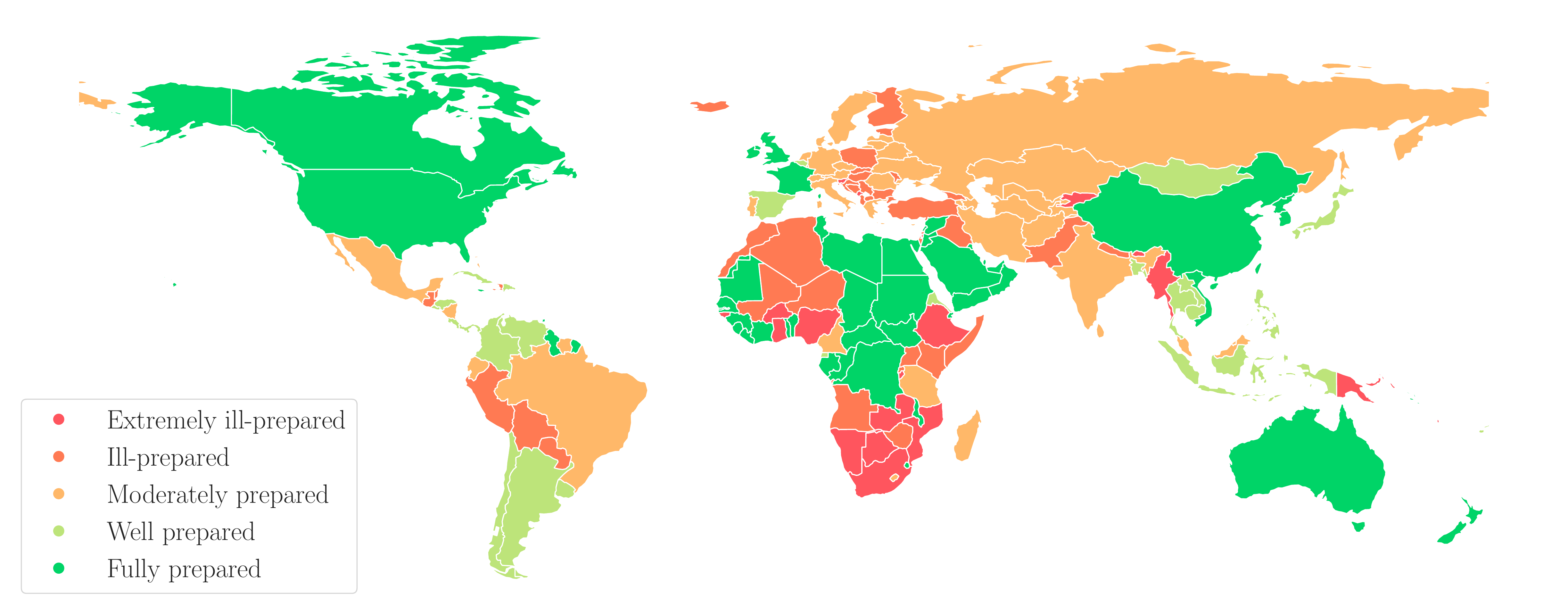}}
    \caption{World map showing the Readiness of each country in terms of combating the next pandemic using LT.}
    \label{fig:readiness}
\end{figure*}

\subsection{Entity Recognition}
\label{sec:entity}
We also evaluate the developed chatbots on another core task of NLU, i.e. entity recognition \cite{DBLP:journals/corr/abs-2007-04248} on English, Hindi and Bengali.
To train and evaluate different COVID bots on this task, we use a set of 200 user related queries (obtained by augmenting existing dataset of 147 queries).
Entity types were identified from a subset of labels from the CORD-19 NER dataset \cite{LuWang2020}, and the queries were accordingly tagged by two native speakers of Bengali and Hindi. Overall, our dataset had a mix of \emph{medical} and \emph{non-medical} entities.
\vspace{-1mm}
The final set of \emph{medical} entity types consists of: \texttt{Covid} (COVID-related entities), 
\texttt{PhysicalScience} (technical terms related to  bio-molecular mechanism of the disease), and \texttt{Disease} (any form of illness or symptoms).
The \emph{non-medical} entity types are: \texttt{BodyPart} (name of the body part), \texttt{Country} (country name), \texttt{Duration} (length in days), \texttt{Protection} (ways to protect against COVID, such as `mask', `gloves'), and \texttt{InfoSource} (source of information).
\texttt{Country} (country name).
For generating the equivalent translations, we manually aligned the entity tags in two languages: Hindi (supported by DialogFlow and LUIS) and Bengali (supported by DialogFlow).
In majority of the cases, we observe that domain-specific entities such as \texttt{incubation, ACE-2 Cells, biochemical assays} are hard to predict by these models on languages other than English. For instance, for \texttt{Covid} entity, we observe significant F1-score drop of
24.6\% for Hindi and 42.9\% for Bengali.
However, for non-medical entities, these models were found to perform comparatively better, e.g., drop in F1-score on \texttt{Country} tag was 5.2\% for Hindi and 8.9\% for Bengali.

\section{Measuring Global Readiness}
\label{sec:readiness}
Although our current work focuses on analyzing pandemic-preparedness of only \numLangs languages, here we try to generalize our findings to other languages by introducing
 a \textit{Readiness Score} for every language which empirically measures the preparedness of current LT to serve its speakers in a pandemic-like emergency situation.
The definition of readiness is based on the assumption that one has access to the best available LT by considering the highest intent detection accuracy of $A_l^{*}$ for a language across different frameworks and training setups.
We then define the readiness of a language $l$ as its relative accuracy with respect to English as $r_l = \frac{A_l^{*}-A_{random}}{A_{en}^{*}-A_{random}}$, where $A_{en}^{*}$ denotes the best case accuracy on English, and 
$A_{random}$ is the accuracy of a random classifier: $A_{random}=100/numberOfIntents$.

We would like to interpolate $r_l$ for all the languages of the world, and hence would need more training examples than the 16 languages that we currently have. We select a set of \numProxyLangs proxy languages (details in Appendix~\ref{appendix:lang_readiness}), ensuring the coverage of major language families in the world\footnote{As per \emph{Ethnologue 24 (2021)} in \url{https://en.wikipedia.org/wiki/List_of_language_families}}.
For these languages, we compute \emph{proxy} accuracies $\tilde{{A}}_{l}^{*}$ by building and evaluating chatbots on MT translated data. We then train a Gaussian Process Regression model for predicting readiness scores with the $r_l$ values for the \numTrainLangs languages as our training set. We use geographical and genetic features from the URIEL database \cite{littell-etal-2017-uriel} 
to represent the languages. The predictive model, which has an average absolute prediction error of 5\%, is then used to estimate the readiness scores of 116 new languages supported by major MMLMs (mBERT and XLM-R) and/or translators (Google and Microsoft).
For all other languages, we set $r_l = 0$, as one can expect near random performance without any LT, as we did see for Kikuyu (Table \ref{tab:results}).

The estimated final $r_l$ scores for all the languages was used to extrapolate the pandemic-readiness of each country $c$, as follows. We use the country-wise language and speaker demographic data\footnote{Infoplease Languages Spoken in Each Country of the World: \url{https://bit.ly/3HoAs9K}} to calculate the country-wise readiness (similar to \citet{blasi2021systematic}),
$r_c = \sum_{l \in \mathcal{L}_c}s_{c,l}r_l$, 
where $\mathcal{L}_c$ is the set of languages spoken in country $c$, and $s_{c,l}$ is the fraction of $c$'s population forming native speakers of the language $l$.

The $r_c$ values  
were clustered to generate five classes ({Extremely ill-prepared:} 0-0.43, {Ill-prepared:} 0.43-0.76, {Moderately prepared:} 0.76-0.86, {Well prepared:} 0.86-0.92, {Fully prepared:} 0.92-1) using Jenks' natural breaks optimization \cite{jenks1967data}. These classes were used to generate a readiness heatmap of the world (Figure \ref{fig:readiness}).

\noindent
\textbf{Observations:}
From Figure \ref{fig:readiness} one can observe that South and East African, and South Asian and East European countries fall into Extremely ill-prepared category, due to the high dominance of low-resource languages in these geographical regions. For instance, people in Zambia's primarily speak Bemba, Chewa and Luzi, all of which are severely under-resourced.
As pointed out in \citet{anastasopoulos-etal-2020-tico}, these regions might also be worse-hit in a pandemic situation, and therefore, require immediate attention.
For Ill-prepared regions such as Bolivia and Paraguay in South America, and Guatemala in Latin America, $r_l$ values are slightly better due to the abundance of Spanish speakers, however there is a sizeable population speaking under-served languages such as Q'eqchi and Guarani.

Countries that fall within fully to moderately prepared  categories typically have large native speaker population of one or more of the class 5 languages (English, French, Chinese, Arabic) and/or well-supported languages (e.g., Korean, Bengali, Malay).
It is important to note that while approximating readiness of a language, we assumed same value for all its diverse linguistic variants and dialects, which in certain cases  results in overestimation of $r_c$.
For instance, high $r_c$ for north and central African countries (e.g., Libya, Egypt and Sudan)
might be due to the sizeable population of a resource-rich language Arabic. However, Arabic has several dialects, which vary from the Modern Standard Arabic at various linguistic levels, and consequently the performance of LT systems for such dialects also vary considerably \cite{zbib-etal-2012-machine, Alsharhan2020}. It holds true for Spanish and Portuguese spoken in Latin America \cite{Lipski_castileand, Lipski+2014+38+75} and French dialects of Western Africa.

\section{Conclusion and Recommendation}
\label{sec:reco}
From our chatbot development experiences, we uncover a set of interesting insights to arrive at the following recommendations which can improve the state of preparedness of languages to combat the next pandemic.

\noindent
\textbf{---} Our experiments showed that low-resource Indian languages (such as Marathi, Bengali) were benefited due to the presence of a geographically and/or linguistically closely related well-resourced language (Hindi). This notion of such ``bridge" languages 
has been explored before in the context of MT \cite{mt_pivots} and zero/few-shot transfer in MMLMs \cite{lauscher-etal-2020-zero}. We recommend the community to target bridge languages for the regions that are currently poorly prepared from an LT perspective.\\

\noindent
\textbf{---} Drawing insights from the brittleness of MT for domain-specific terms (\textit{airborne, incubation}) or newly-coined terms ($\textit{COVID}$), we believe that commercial and open-source bot frameworks can benefit from domain adaptation techniques \cite{chu-wang-2018-survey}, or techniques to inject new terms to existing solutions.

Our study confirms that except English, only a few European and Asian languages push forward the state-of-the-art research in LT for healthcare.
Our preliminary investigation suggests that instead of demographic demand, it is the economic prowess of the users of a language that drives the investment towards developing sophisticated LT solutions for a given language. 
For instance, Swahili, even though considered as the \textit{lingua franca} of Africa, is still under-served by commercial chatbot frameworks. 
Similar trends were observed for Hausa which has a considerably large speaker base compared to Dutch (resource-rich)\footnote{\url{https://en.wikipedia.org/wiki/List_of_languages_by_number_of_native_speakers}}.

We believe that these findings will play a crucial role in making the community aware of the disparity that needs to be addressed before the next pandemic hits.

\bibliographystyle{acl_natbib}
\bibliography{anthology,acl2021}

\section{Example Appendix}
\label{sec:appendix}

\subsection{Intent Definitions and Descriptions}
\label{appendix:intentDesc}
The different intents used for our experiments are described in table ~\ref{table-intentDesc}. We provide definitions and examples for each of the different intents used.

\begin{table*}[!t]
\footnotesize
\begin{center}
 \begin{tabular}{p{2.7cm} | p{5.4cm} | p{6.8cm} } 
 \hline
 \textbf{Intent Type} & \textbf{Example in English} & \textbf{Definition}\ \\
 \hline
\texttt{Airborne} & Can the virus that causes COVID-19 be transmitted through the air? & Queries related to how much COVID is carried by air \\
 \texttt{ClarifyCovid} & How do I know if it is COVID-19 or just the flu? & Queries related to difference betwen COVID and other diseases \\
 \texttt{Incubation} & Can someone in incubation infect other people? & Queries related to situations where a person is infected with COVID and is going through incubation phase \\
 \texttt{Length} & How long does the illness make you poorly for? & Queries regarding longevity of COVID infection \\
 \texttt{Mask} &  Should I wear a mask while exercising? & Queries about wearing mask  \\
\texttt{Protection} & Ways to keep safe from COVID-19 & Queries about the ways of protection from  COVID \\
\texttt{Quarantine} & Will I avoid coronavirus, if I self-isolate? & Queries about the effect of quarantining after getting infected with COVID \\
 \texttt{Spread} & Aside from inhalation, are there other ways coronavirus can spread? & Queries about the spreading process of COVID\\
 \texttt{Testing} & Where can I get my test done? & Queries about the testing process of COVID\\
 \texttt{CovidTwice} & If you get COVID-19, can you get it again? & Queries about whether COVID can infect someone more than once \\
\texttt{ExplainSymptom} &  I have a sharp pain here in the chest & User explaining COVID related symptoms \\ 
\texttt{Country} & How many people in Italy have COVID-19? & Querying about the statistics of infection in different countries \\
\texttt{Medication} & Do any of the drugs reduce mortality? & Querying about the medication to survive from COVID \\
\texttt{Treatment} & Which vaccines are good to protect against the virus? & Querying about the treatment strategies associated with COVID \\
\bottomrule
\end{tabular}
\end{center}
\caption{\label{table-intentDesc}Different intents with definitions and examples present in our dataset.}
\end{table*}

\subsection{Bot Building Strategies}
\label{appendix:bot}

\begin{table*}[!t]
\footnotesize
\begin{center}
 \begin{tabular}{p{2.7cm} | p{5.4cm} | p{6.8cm} } 
 \hline
 \textbf{Bot Building Setup} & \textbf{Training Strategy} & \textbf{Testing Strategy}\ \\
 \hline
\texttt{Train on English Data} & Train set comprises of the English queries & Test set comprises of English queries where the manually written queries in target language are translated to English using MT system \\
\texttt{Train on MT Translations} & Train set comprises of the English queries translated to target language using MT System & Test set comprises of manually written queries in target language \\
 \texttt{Train on Manual Translations} & Train set comprises of manually written queries in target language & Test set comprises of manually written queries in target language \\
\bottomrule
\end{tabular}
\end{center}
\caption{\label{table-botsetup} Different strategies for building the chatbots.}
\end{table*}

\begin{figure*}
     \centering
     \includegraphics[width=0.9\textwidth]{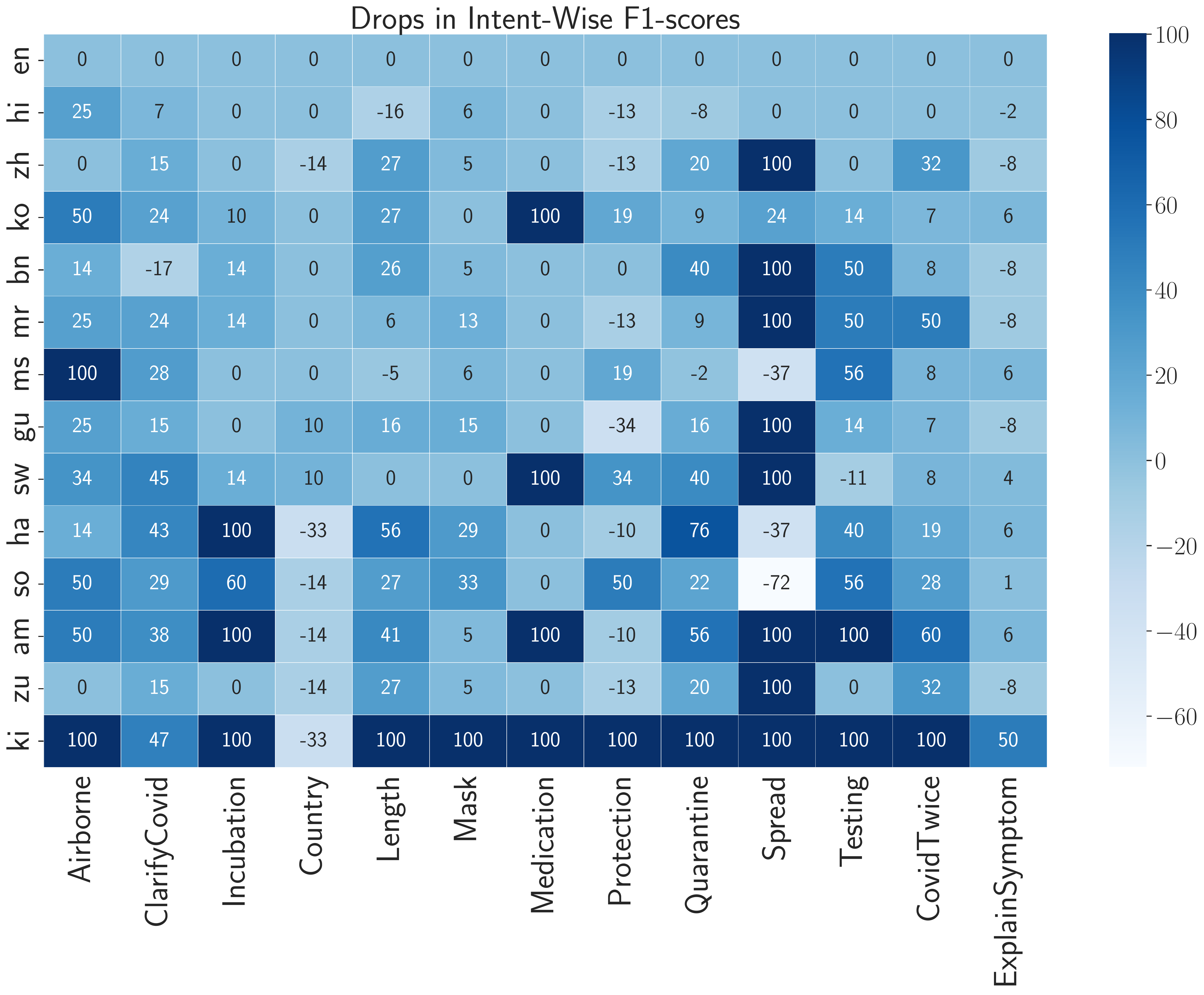}
     \caption{Relative drops (relative to English) in intent wise F1 scores for different languages in the \textit{Train on English} setup (in LUIS). Negative values indicate increase in the scores relative to English.} 
     \label{fig:intent_f1_tt}
\end{figure*}

\begin{table*}[t]
\footnotesize
\centering
\begin{tabular}{l l ccc ccccc}
\toprule
&& \multicolumn{3}{c}{Medical} & \multicolumn{5}{c}{Non-Medical} \\
\cmidrule(r){3-5}
\cmidrule(r){6-10}
Lang & Bot & Covid & PhysicalScience & Disease & BodyPart & Country & Duration & Protection & InfoSource\\
\midrule
\multirow{2}{2em}{Hi} & DF & 24.6
 & 50.1
 & +3 & 32 & 7.52 & 5.2 & 11.3 & +30.1 \\
 & LUIS &  24.56
 & +41
 & 94 & 21 & 5.31 & +4 & 12.06 & +72.72\\
 \cmidrule(r){1-1}
\multirow{1}{2em}{Bn} & DF & 42.9
 & 43.1
 & 52 & 34 & 6.3 & 8.9 & 20.4 & 8.34 \\
\bottomrule
\end{tabular}
\caption{Relative drop in entity-type wise F1-score in \textit{Entity Recognition} task using DialogFlow (DF) and LUIS.}
\label{tab:entity_results}
\end{table*}

\subsection{MMLM Training Setup}
\label{appendix:mmlm}
For our experiments with Multilingual Pre-trained Transformers we consider mBERT (\textit{bert-base-multilingual-cased}) and XLMR (\textit{xlm-roberta-base}) for training intent classifiers. As mentioned in the main text we explore two methodologies to train and evaluate these MMLMs, a detailed description with hyperparameters is given below: \\

\noindent
\textbf{1. KNN using Pre-trained Embeddings}: \\
Since the scale of our data is on the lower side, training an end-to-end classifier might be prone to over-fitting. We fit a k-Nearest Neighbors (KNN) classifier on the sentence embeddings obtained using the pre-trained model for the queries in training data. At test time, we similarly obtain the representation of the user query and find its nearest neighbors among the training queries to predict its intent. The optimal value for $k$ was empirically found to be 1 and for sentence embeddings, we take the average of the representation of each token of the sentence in the last layer of MMLM.

We also tried fine-tuning the pre-trained model with the training queries using a Masked Language Modelling (MLM) objective. Additionally, we also fine-tuning on a much larger COVID-19 queries dataset in english : COQB \cite{li-etal-2020-jennifer} along with our training queries, as has been pointed by \citet{lauscher-etal-2020-zero} can be an effective strategy for few shot transfer. We use 3 epochs to fine-tune the models with  a learning rate of $5\text{e-}5$ and Adam-W optimizer \cite{loshchilov2018decoupled}. A masking probability of 15\% was used during the MLM training and maximum sequence length was taken to be 32. \\

\noindent
\textbf{2. Fine-tuning an End to End Classifier} : \\
We also try fine-tuning the MMLMs end-to-end by adding a classification head on top of the pre-trained network to classify the input query into one of the \numIntents intents. We adapt the sequence classification scripts for GLUE benchmark \cite{wang-etal-2018-glue} provided by hugging face\footnote{\url{https://huggingface.co/transformers/v2.9.1/examples.html}} on our dataset. We fine-tune the classifier for 20 epochs, with the same learning rate and optimizer as the MLM fine-tuning in the first point with a batch size of 8.

For every language we use the best accuracies obtain from either of these two strategies \footnote{technically 4, as in the KNN case we consider no fine-tuning, fine-tuning on Train Queries, and fine-tuning on Train and COQB queries}. All the experiments were run on 4 NVIDIA V-100 GPUs with 32 GB memory.

\subsection{Language Readiness Analysis}
\begin{figure*}[!t]
    \centering
    \fbox{\includegraphics[width=0.96\textwidth]{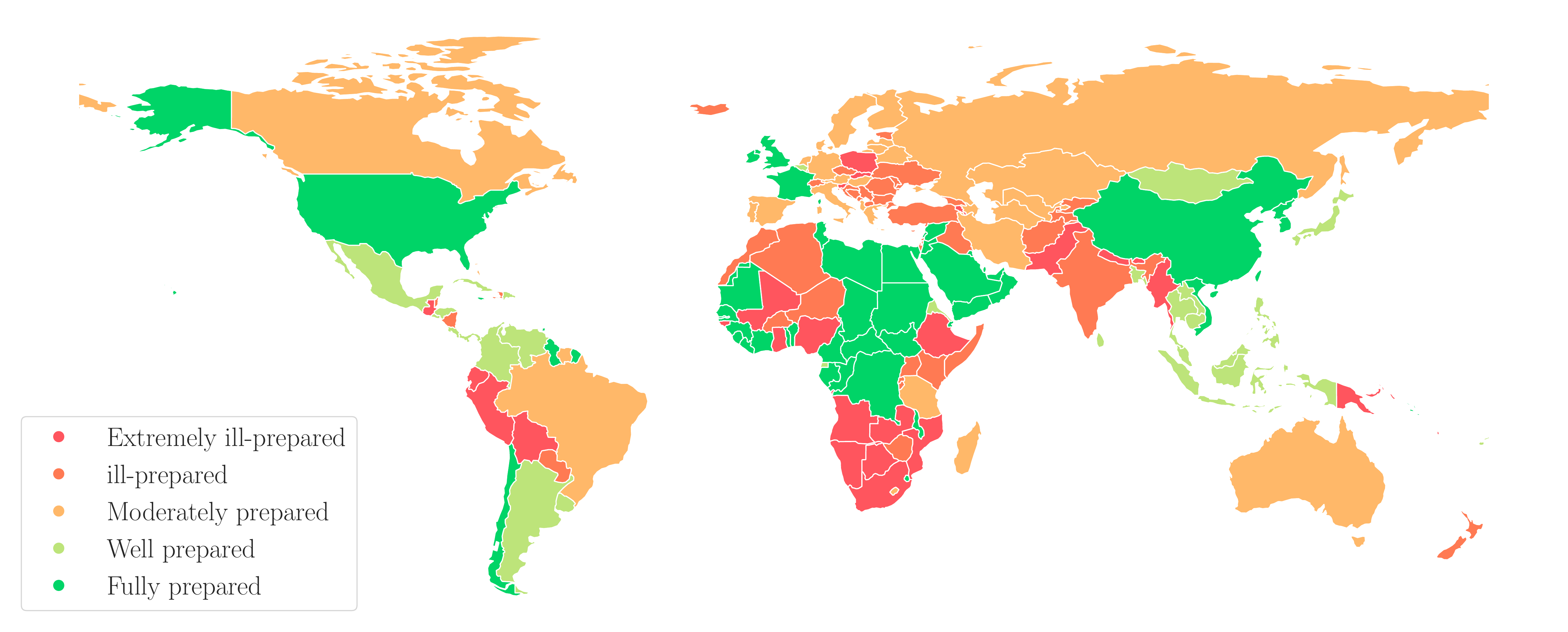}}
    \caption{World Map showing the Readiness of each country in terms of combating the next pandemic using LT. Their Levels of Preparedness are shown as legends in the bottom left corner. This map was generated by providing uniform weightage to all the languages spoken in a country, i.e. excluding the percentage of speaker-base for a particular language in a country (Readiness measure calculated using Equation 2).}
    \label{fig:scplot_language}
\end{figure*}

\noindent
\textbf{Results and Analysis} \\
\noindent
Initially, we have plotted the readiness measures of each language used in our training data on the scatter plot in Figure ~\ref{fig:scplot_language} with the language class on X-axis and readiness measure in Y-Axis. It clearly shows that the African languages such as  \texttt{Somali, Amharic, Hausa, Zulu} are below the trendline in terms of readiness. In fact, some of the European languages such as  \texttt{Icelandic, Hungarian, Estonic, Finnish} also require some attention. Primarily, we observe that the readiness measure is not a direct function of the language class from this plot. As we can see that even though majority of the class 4 and 5 are near the trend line, the observation is similar for Class 1 as well. 

\begin{figure}
    \centering
    \includegraphics[width=0.47\textwidth]{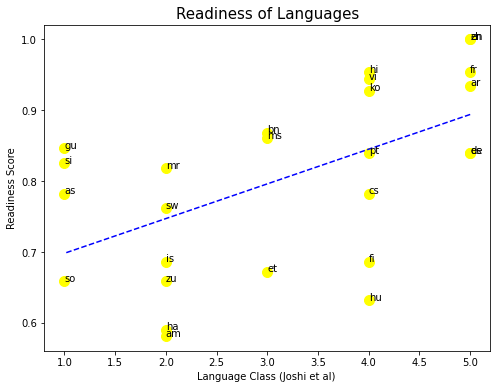}
    \caption{shows the readiness scores of the languages which are used in our training data for readiness measurement using GPR}
    \label{fig:scplot_language_family}
\end{figure}

Therefore, we also resort to understanding how much does the trend hold true for the language families of these corresponding languages? So, we approximate each of the language family by taking the average scores of each language falling into that class and plot those in Figure ~\ref{fig:scplot_language_family}. It was interesting to observe that the English-major language families such as \texttt{Austroasiatic, Koreanic} and \texttt{Sino-Tibetan} are well-served, and consequently lie above the trend line. Overall, \texttt{Indo-European} language families are well near the trend line and then the resource-poor language families are \texttt{Afroasiatic, Niger-Congo} and \texttt{Uralic}, the worst being the \texttt{Afroasiatic} language family.\\

\subsection{Details on Language Readiness Prediction}
\label{appendix:lang_readiness}
In section \ref{sec:readiness}, we discussed the estimation of readiness values of different languages. We first extended our \numLangs languages that we considered for intent recognition experiments with proxy scores for an additional \numProxyLangs languages, namely, \texttt{French (fr)} , \texttt{Arabic (ar)}, \texttt{German (de)}, \texttt{Spanish (es)}, \texttt{Portuguese (pr)}, \texttt{Vietnamese (vi)}, \textit{Hungarian (hu)}, \texttt{Finnish (fi)}, \texttt{Czech (cs)}, \texttt{Estonian (et)}, and \texttt{Icelandic (is)}. Finally, it covers six primary language families in the world, such as: 1) \texttt{Indo-European}, 2) \texttt{Sino-Tibetan}, 3) \texttt{Afroasiatic}, 4) \texttt{Niger-Congo}, 5)  \texttt{Koreanic} and 6) \texttt{Austroasiatic}.
To estimate the readiness values of the remaining 116 languages supported by the Translators (Google and Bing) and MMLMs (mBERT and XLMR), we used the available readiness data for the \numTrainLangs to build a regression model. We used Gaussian Processes to model the readiness prediction problem, due its efficiency on the small sized datasets. Radial Basis Function (RBF) with added noise level for each instance (White Kernel), was used, and the \textit{length scale} of RBF and noise level were tuned using L-BFGS algorithm with 5 restarts for the optimizer. The model selection was done using a Leave One Out strategy, where we move one language to validation set and train on the remaining, repeating this for all the languages and measuring average accuracy. Besides Gaussian Process Regression (GPR), we also experimented with Linear Regression, Lasso Regression and XGBoost \cite{xgboost}, but observed inferior validation accuracies.

\begin{figure}
    \centering
    \includegraphics[width=0.47\textwidth]{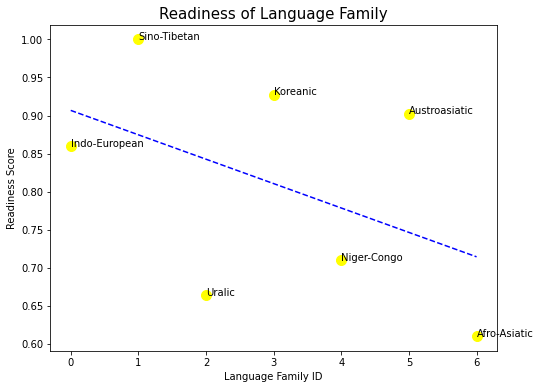}
    \caption{shows the readiness scores of the language families  of the corresponding languages which are used in our training data for readiness measurement using GPR}
    \label{fig:entity_eg1}
\end{figure}

\subsection{Global Pandemic Readiness Measurement}
\label{appendix:country_readiness}
In section 4 of the paper, we have talked about how to actually take the speaker-base values into account while calculating the readiness scores for each country in the world and the final $r_l$ scores obtained on all the languages are used to extrapolate the readiness of each country $c$ in the world.
We had also experimented in a way such that all the languages spoken in the country are weighted equally while calculating the readiness of a country. This is similar to the linguistic utility defined by \citet{blasi2021systematic} in their work
for a country $c$ we calculate linguistic readiness $r_c^{ling}$ as:
\begin{equation}
    r_c^{ling} = \frac{1}{|\mathcal{L}_c|} \sum_{l \in \mathcal{L}_c}r_l
\end{equation}
The $r_{c}^{ling}$ values have been plotted in Figure \ref{fig:scplot_language}. Based on our observations on these values we make the following observations highlighting the difference between demographic and linguistic readiness of different countries.

\noindent
\textbf{Observations:}
The map shown in ~\ref{fig:entity_eg1} provides us an idea of how each country in the world would be able to effectively combat the pandemic by leveraging LT solutions.
However, this is when we are actually considering uniform speaker-base for each language in a country.
Overall, it can be observed that some of the Asian countries like India falls in the  \textit{\textbf{ill-prepared}} zone now which was initially treated as \textit{\textbf{moderately prepared}}. 
This is due to the presence of class 4 language Hindi (having a readiness score of 0.9536) with a considerably high speaker-base (46.19\%). 
Also, similar trend is observed in Canada (home to the speakers of various languages like \texttt{English, French, Punjabi, Italian, Spanish,} \texttt{German, Cantonese, Arabic, Tagalog}).

\end{document}